\documentclass[10pt,conference]{IEEEtran}

\usepackage[utf8]{inputenc}
\usepackage[T1]{fontenc}
\usepackage{cite}
\usepackage{amsmath,amssymb,amsfonts}
\usepackage{graphicx}
\usepackage{booktabs}
\usepackage{balance}
\usepackage{tabularx}
\usepackage{multirow}
\usepackage{makecell}
\usepackage{enumitem}
\usepackage{xcolor}
\usepackage{xspace}
\usepackage{url}
\usepackage[hidelinks]{hyperref}
\usepackage{ifthen}
\usepackage{pifont}
\usepackage[strict]{changepage}
\usepackage{framed}
\usepackage{siunitx}
\usepackage{adjustbox}
\newcommand{\wcircle}[1]{\ding{\numexpr171 + #1}}
\newcommand{\bcircle}[1]{\ding{\numexpr181 + #1}}

\newcommand{\mz}{\textsc{MiniZoo}\xspace}

\newcommand{\mzSnapshotDate}{November 9$^{th}$ 2026\xspace}

\newcommand{\mzAppsWithPackages}{\num{13009}\xspace}
\newcommand{\mzDistinctPackages}{\num{16273}\xspace}
\newcommand{\mzPackageVolumeGiB}{\num{30.33}\xspace}

\newcommand{\mzAppsWithMultipleVersions}{\num{1149}\xspace}

\newcommand{\mulversionpct}{\num{8.83}\%\xspace}

\usepackage{CJKutf8}
\newcommand{\zh}[1]{\begin{CJK*}{UTF8}{gbsn}#1\end{CJK*}}

\definecolor{promptbg}{HTML}{E6F3FF}
\definecolor{promptborder}{HTML}{2166A5}
\definecolor{answerbg}{RGB}{225,255,225}
\definecolor{answerborder}{RGB}{0,140,75}

\begin{document}

\title{MiniZoo: An Evolving Dataset of Mini Apps for the Research Community}

\author{\IEEEauthorblockN{Luca Ferrari}
\IEEEauthorblockA{\textit{IMT School for Advanced Studies Lucca}\\Lucca, Italy\\luca.ferrari@imtlucca.it}
\and
\IEEEauthorblockN{Marco Alecci}
\IEEEauthorblockA{\textit{SnT, University of Luxembourg}\\Luxembourg, Luxembourg\\marco.alecci@uni.lu}
\and
\IEEEauthorblockN{Jordan Samhi}
\IEEEauthorblockA{\textit{SnT, University of Luxembourg}\\Luxembourg, Luxembourg\\jordan.samhi@uni.lu}
\and
\IEEEauthorblockN{Tegawend\'e F. Bissyand\'e}
\IEEEauthorblockA{\textit{SnT, University of Luxembourg}\\Luxembourg, Luxembourg\\tegawend\'e.bissyande@uni.lu}
\and
\IEEEauthorblockN{Jacques Klein}
\IEEEauthorblockA{\textit{SnT, University of Luxembourg}\\Luxembourg, Luxembourg\\jacques.klein@uni.lu}
\and
\IEEEauthorblockN{Luca Verderame}
\IEEEauthorblockA{\textit{University of Genoa}\\Genova, Italy\\luca.verderame@unige.it}}

\maketitle

\begin{abstract}
  Mini apps run within host applications (apps) such as WeChat, Alipay, Baidu, and Telegram, providing services without requiring users to install separate native apps. Their growing ecosystem has attracted research on software quality, security, and privacy, creating a need for reusable datasets. However, building and maintaining such datasets is challenging, requiring platform-specific collection infrastructure and mechanisms to handle incomplete packages, changing availability, and heterogeneous metadata.
A shared, continuously maintained collection is still lacking, therefore motivating the development of a common research resource, whose value has been demonstrated by the experience of AndroZoo in the Android ecosystem.

In this paper, we present \mz, an evolving corpus of mini apps and associated metadata. The initial release comprises \mzDistinctPackages{} mini-app packages, with multiple versions of the same mini apps. 
The dataset is managed and distributed through AndroZoo under its existing access conditions. To facilitate its reuse, we describe the collection process, data organization, provenance, and limitations, and explain how researchers can select and access the data.

\mz is designed to grow continuously through the addition of newly discovered mini apps, the inclusion of further host ecosystems, and the collection of successive versions of previously collected apps. This ongoing expansion will enable both comparisons across ecosystems and longitudinal studies of how mini apps evolve over time.

\end{abstract}

\begin{IEEEkeywords}
  Mini Apps, WeChat, Software Repository, Dataset, Software Evolution
\end{IEEEkeywords}

\section{Introduction}
\label{sec:introduction}

Mini apps are lightweight applications (apps) that run within a host app, often called a \emph{super app}~\cite{Schreieck2023}. WeChat~\cite{tencentminiprograms}, Alipay~\cite{alipayminiprograms}, Baidu~\cite{baidusmartprograms}, and Telegram~\cite{telegramminiapps} are among the platforms that support mini apps, exposing APIs through which these apps interact with users and access host-provided services. As mini apps expose substantial client-side functionality through platform-specific code, configuration, and interface resources, they have attracted growing interest in software engineering, security, and privacy research. Existing work has investigated mini-app bugs~\cite{wang2022webug}, sensitive data flows~\cite{wang2023taintmini,meng2023wemint}, privacy-policy consistency~\cite{wang2024minidetector,wang2025miniscope,ferrari2026telegapper}, and credential leakage across multiple platforms~\cite{shi2025keymagnet}, among other topics~\cite{zhang2021measurement,li2024xposcope,li2024minitracker,wang2023onesize,yang2022cmrf,wang2023apiscope}.

Conducting such research requires access to the apps themselves, yet collecting mini apps involves more than downloading files from a public web catalogue: researchers must discover apps, interact with host platforms, and retrieve packages together with associated metadata.
Existing studies do not always release their collected packages, with public artifacts often focusing on the code for crawlers or analysis tools~\cite{zhang2021measurement,wang2025miniscope,zhang2024minicat,shi2025keymagnet}. Other initiatives, despite reporting collections containing more than one million mini apps, only share selected samples or datasets tailored to particular research questions~\cite{yang2025malware,minisec2026}. Publicly downloadable artifacts often contain extremely small datasets (e.g., 10 apps)~\cite{shi2026cloud}. Additionally, even when large, these collections provide only a fixed snapshot of the mini apps without preserving successive versions, limiting their usefulness for longitudinal or evolution studies.
This research gap highlights the need for a large, shared, continuously maintained and evolving collection of mini apps. A successful example from which we draw inspiration is AndroZoo, a continuously growing collection of Android apps ($\sim$28M apps in November 2026) used by thousands of researchers (more than 1600 citations)~\cite{allix2016androzoo,alecci2024retrospective}.

In this paper, we present \mz, an evolving dataset of mini apps for the research community. To build and maintain this resource, we have established automated crawling infrastructure to discover mini apps and collect their packages, associated metadata, and acquisition records. Although the current corpus focuses on WeChat, \mz follows a platform-agnostic organization that we plan to extend to additional mini-app ecosystems through dedicated crawlers while retaining a common data structure and access framework. As of \mzSnapshotDate{}, \mz contains \mzAppsWithPackages{} mini apps, of which \mzAppsWithMultipleVersions{} (\mulversionpct{}) already have multiple recorded versions, for a total of \mzDistinctPackages{} distinct packages.
These packages and metadata support research on software quality, security, and privacy, while continued collection will expand opportunities for longitudinal studies of mini-app evolution. \mz is hosted and managed through AndroZoo~\cite{allix2016androzoo,alecci2024retrospective}, allowing us to reuse its existing infrastructure, draw on its maintenance expertise, and reach its established research community. The dataset and the access conditions are available at:
\begin{center}
\url{https://androzoo.uni.lu/minizoo}
\end{center}
\noindent
\section{Dataset Overview}
\label{sec:dataset}

\mz is designed as a continuously growing collection of mini-app packages and metadata for the research community. Each mini-app platform has its own architecture and mechanisms for mini-app discovery and package retrieval, requiring dedicated crawlers whose development and maintenance are challenging and time-consuming. We therefore selected WeChat as our initial target because it has been extensively investigated in the research literature~\cite{zhang2021measurement,wang2023taintmini,wang2025miniscope}, with concrete plans to extend the collection to other mini-app platforms. Mini-app platforms can also differ in their package formats and execution environments; for example, WeChat uses the \texttt{.wxapkg} format (see Section~\ref{sec:collection}) and provides a runtime within the host app to load and execute mini-app code. Although the current corpus focuses on WeChat Mini Programs, the dataset follows a platform-agnostic organization to support planned extensions to other ecosystems, such as Alipay, Baidu, and Telegram. Its core abstractions and metadata schema provide a common structure that can be extended with platform-specific fields as needed.

\noindent
\textbf{Definitions: Mini Apps and Packages.}
An important distinction must be made between \wcircle{1} a mini app and \wcircle{2} its packages. A mini app is identified within its host platform by a unique identifier, such as the AppID in WeChat, which allows different versions of the same mini app to be associated with a single identity. A package is an archive containing client-side code and resources that can be identified by a hash (e.g., SHA-256). If an update changes the archive’s contents, the resulting package has a different hash while remaining associated with the same mini app. For example, if a weather mini app receives such an update, its AppID would remain the same, while the SHA-256 hash identifying its package would differ from that of the previous package. This distinction is analogous to that between Android apps, identified by package names such as \textit{com.android.chrome}, and APK files, the archives containing their code and resources. In AndroZoo, apps are identified by their package name, while individual APK files are identified by their SHA-256 hashes, with multiple APKs sharing the same package name and representing different versions of the same app~\cite{allix2016androzoo,alecci2024retrospective}.

\noindent
\textbf{Dataset Content.}
For WeChat, we decided to rely on AppIDs to identify mini apps and on SHA-256 hashes to identify their packages, distributed in the platform-specific \texttt{.wxapkg} format described later in Section~\ref{sec:collection}. Table~\ref{tab:statistics} summarizes the corpus as of \mzSnapshotDate{}, while updated statistics are available on the \mz website.
 
\begin{table}[ht]
\caption{Corpus overview as of \mzSnapshotDate. }
\label{tab:statistics}
\centering

\begin{adjustbox}{width=0.8\columnwidth,center}
\begin{tabularx}{\columnwidth}{@{}Xr@{}}
\toprule
\textbf{Property} & \textbf{Value} \\
\midrule
Mini apps with at least one associated package
    & \mzAppsWithPackages \\
Distinct package contents (SHA-256)
    & \mzDistinctPackages \\
Mini apps with multiple recorded versions
    & \mzAppsWithMultipleVersions \\
Total package volume (GiB)
    & \mzPackageVolumeGiB \\

\bottomrule
\end{tabularx}
\end{adjustbox}

\end{table}

The statistics describe the cumulative corpus collected from September to \mzSnapshotDate{}, based on the consolidated export generated on that date. The corpus contains \mzAppsWithPackages{} mini apps with at least one associated package, comprising \mzDistinctPackages{} distinct package contents and a total volume of \mzPackageVolumeGiB{} GiB. Package counts and volume are deduplicated by SHA-256, so repeated acquisitions of identical files do not inflate these figures. Of the collected mini apps, \mzAppsWithMultipleVersions{} (\mulversionpct{}) have multiple recorded versions, providing an initial basis for studying changes within the same apps. Continued collection will expand both the number of represented apps and the version histories available for longitudinal analysis.

\noindent
\textbf{Data Organization.}
Users have access to a single compressed \texttt{tar.gz} archive, which download instructions are provided later in Section~\ref{sec:access}. The archive comprises three main components:

\begin{enumerate}[leftmargin=*]
    \item \texttt{minizoo.csv}: a CSV file listing all the mini apps available in \mz and their associated packages.
    \item The acquired packages, i.e., \texttt{.wxapkg} files.
    \item \texttt{minizoo\_metadata.jsonl}: a JSON Lines file containing structured metadata complementing the CSV list with additional information that researchers can use.
\end{enumerate}

While this initial distribution model provides access to the complete corpus in a single download, we plan to extend the API to support selective package retrieval as the collection grows. Researchers will then be able to specify the SHA-256 hashes of the packages they need or to download all the packages related to a specific appID, following AndroZoo’s existing download model and avoiding the need to download the entire archive for studies focusing on a subset of the collection. 

In addition to the packages themselves, \mz provides metadata describing the collected mini apps, their packages, and acquisition times, helping researchers interpret the corpus and select data relevant to their studies. Table~\ref{tab:metadata} summarizes the metadata provided through the CSV and JSON Lines files.

\begin{table}[ht]
\caption{Overview of the metadata provided by \mz.}
\label{tab:metadata}
\centering
\small
\begin{adjustbox}{width=0.8\columnwidth,center}
\begin{tabularx}{\columnwidth}{@{}p{0.25\columnwidth}X@{}}
\toprule
\textbf{Group} & \textbf{Representative information} \\
\midrule
Identity & AppID, SHA-256 hash\\
Mini-app related & name, description, category, developer, icon URL \\
Package related & size, version code, collection time \\

\bottomrule
\end{tabularx}
\end{adjustbox}
\end{table}

The \texttt{minizoo.csv} file contains the fields needed to identify and select packages:
\begin{itemize}[leftmargin=*]
    \small
    \item \textbf{AppID}: The unique identifier assigned to the mini app.
    \item \textbf{SHA-256}: The hash identifying the acquired \texttt{.wxapkg} file.
    \item \textbf{Package size}: The size of the acquired package file.
    \item \textbf{Version code}: The version code associated with the mini app.
    \item \textbf{Collection time}: The timestamp of when the package was collected.
\end{itemize}

The \texttt{minizoo\_metadata.jsonl} file provides descriptive information about each mini app: 
\begin{itemize}[leftmargin=*]
    \small
    \item \textbf{Name}: The name presented to the users.
    \item \textbf{Description}: The textual description of the mini app.
    \item \textbf{Category}: The category associated with the mini app.
    \item \textbf{Developer}: Information identifying the developer or organization responsible for the mini app.
    \item \textbf{Icon URL}: The URL pointing to the mini app’s icon.
\end{itemize}
Each JSON Lines record also includes the AppID, allowing researchers to link these app-level metadata to the corresponding package entries in the CSV list.
\section{Data Collection and Maintenance}
\label{sec:collection}

\begin{figure}[ht]
\centering
\includegraphics[width=0.9\columnwidth]{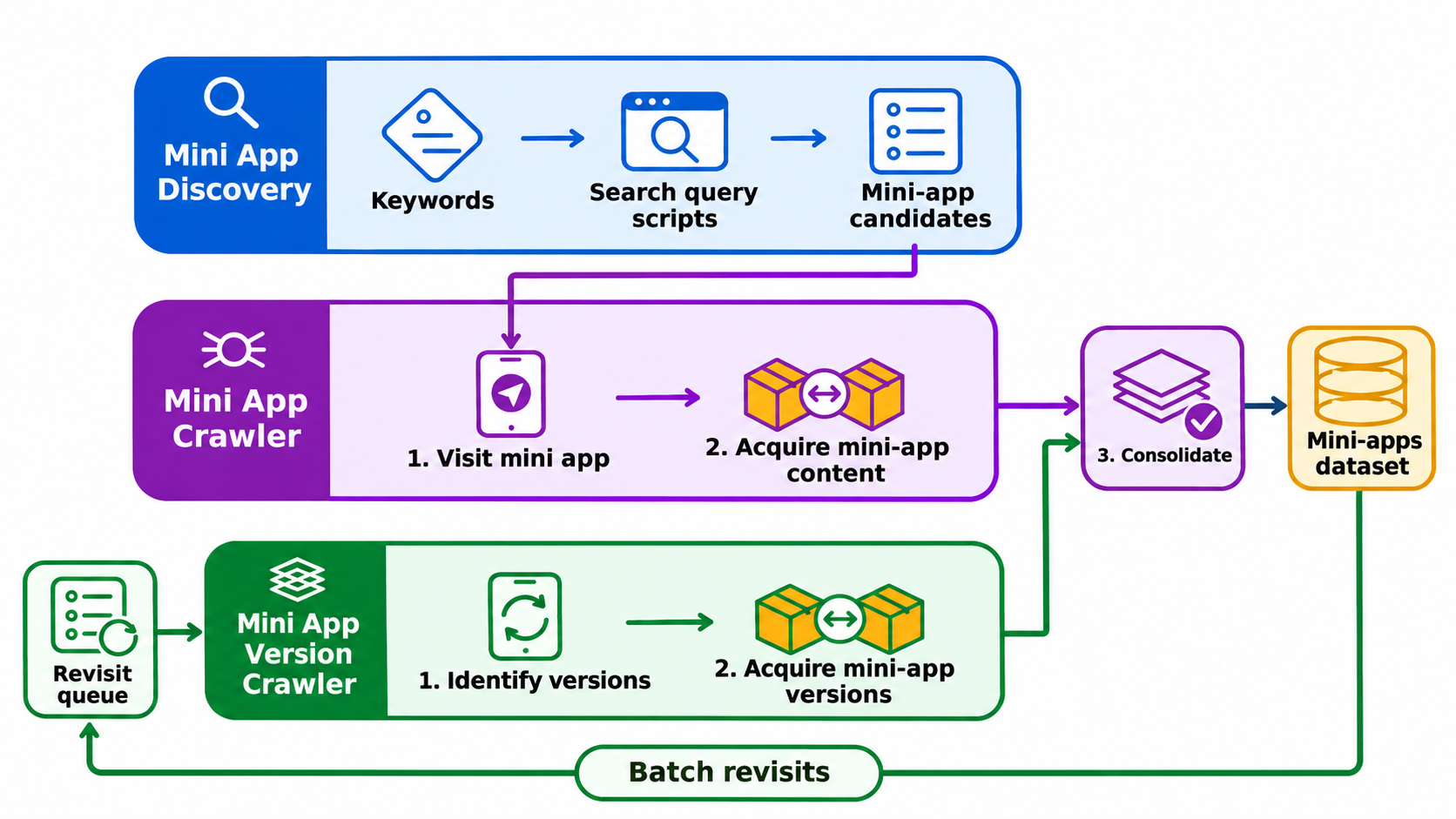}
\caption{Collection workflow with initial and version crawlers.}
\label{fig:approach}
\end{figure}

Our collection workflow is designed to expand \mz{} with newly discovered mini apps while at the same time adding updated packages of already collected mini apps. As illustrated in Figure~\ref{fig:approach}, the workflow comprises three complementary modules: \bcircle{1} \textit{Mini App Discovery}, which identifies mini apps through keyword-based searches; \bcircle{2} \textit{Mini App Crawler}, which acquires packages from newly discovered apps; and \bcircle{3} \textit{Mini App Version Crawler}, which revisits previously collected apps to identify and acquire newer versions. The outputs of both crawlers are then consolidated into a unified corpus that preserves the version history.

\noindent
We describe each module in detail hereafter:

\noindent
\bcircle{1} \textbf{Mini App Discovery.} The goal of this module is to build and expand a list of mini apps whose packages will be downloaded by the crawlers. To this end, a custom script inspired by MiniCrawler~\cite{zhang2021measurement} submits keyword searches to WeChat through Frida~\cite{frida}, storing the returned AppIDs and metadata in a SQLite database while expanding the search space by adding returned app names and overlapping Chinese substrings of two to four characters to the query queue. These results are then combined with existing catalogue and collection records to build the discovery catalogue, which provides the crawlers with the list of apps to collect. The retained discovery log contains 688 distinct queries, including \zh{工具} (tools), \zh{教育} (education), \zh{旅游} (travel), \texttt{weather}, \texttt{CRM}, and app names. The module can run continuously to expand the list of discovered mini apps; however, when a substantial backlog of apps awaiting collection accumulates, discovery can be temporarily paused to prioritize package acquisition and resume once the backlog has been reduced.

\noindent
\bcircle{2} \textbf{Mini App Crawler.} The Mini App Crawler acquires packages from newly discovered apps, with collection coordinated by a supervisor, i.e., a process that distributes pending apps across available devices and prioritizes retries of failed attempts. For each target app, the crawler searches WeChat by name and checks the AppID of the app that opens to verify whether it matches the requested one. If a different app opens, the crawler can still acquire its packages and associate them with its actual AppID, while keeping the originally requested app pending for a subsequent attempt.

\noindent
\bcircle{3} \textbf{Mini App Version Crawler.} The Mini App Version Crawler revisits previously collected mini apps in batches to check for newer versions and acquire their packages, without a fixed revisit interval for individual apps. For each app, it searches WeChat by name and proceeds only if the opened app’s AppID matches the requested one, ensuring that newly acquired packages are compared with those previously collected from the same app. A newer observed version becomes the reference for detecting subsequent updates only after identity, file integrity, completeness, and decompilation checks have passed.

WeChat distributes mini apps in the platform-specific \texttt{.wxapkg} format, which bundles client-side program logic, configuration data, and assets such as images, together with JavaScript code and resources derived from WXML markup and WXSS stylesheets that define the interface structure and appearance. Unpacking and decompiling these packages enables researchers to inspect their contents and perform static analysis~\cite{wang2023taintmini}. To acquire these files, both crawlers copy the main package and all available subpackages linked to the opened app by WeChat's runtime metadata, without clearing WeChat's cache, then verify the copies against WeChat's checksums and inspect the app again to confirm that its identity, version, and package inventory have remained unchanged during acquisition. Before opening the next app, they also store the opened app's available metadata separately from the search input and extract package roles, names, developer information, version strings, and AppID references, where available. The outputs of both crawlers are merged after checking file hashes and the evidence linking packages to AppIDs, using the AppID--SHA-256 pair for deduplication so that separate associations are preserved when collection evidence links identical package contents to different apps.

\noindent
\textbf{Implementation.}
Both crawlers are implemented in Python and use Android Debug Bridge (ADB) for device access and Frida~\cite{frida} to inspect running mini apps, with search navigation relying on ADB and ADBKeyboard~\cite{adbkeyboard} in the Mini App Crawler and on Appium~\cite{appium} in the Mini App Version Crawler. The reported collection was conducted in Genoa, Italy, and Luxembourg, using three rooted devices (one Pixel 9 and two Pixel 9a units), all running Android 16 (API level 36) and WeChat 8.0.72, with the \texttt{en-US} locale and the \texttt{Europe/Luxembourg} time zone. We plan to evaluate emulators to scale collection beyond the available physical devices.
The code associated with our crawlers can be found at: \url{https://github.com/Mobile-IoT-Security-Lab/MiniZoo-Scraper}
\section{Hosting and Access to the data}
\label{sec:access}

\mz is managed through AndroZoo and hosted on the same servers as its Android app collection~\cite{allix2016androzoo}. This integration allows us to draw on the expertise gained from maintaining AndroZoo, reuse its storage and distribution systems, and promote \mz within its established research community.

Access to \mz requires a personal AndroZoo API key, issued under AndroZoo’s existing access conditions (\url{https://androzoo.uni.lu/access}). To request a key, researchers must email \href{mailto:androzoo@uni.lu}{\nolinkurl{androzoo@uni.lu}} from their institutional address, providing their full name, research institution, and institutional email address. A faculty member or another person in a permanent position must agree to these conditions, which restrict redistribution and commercial use, prohibit marketplace creation, and require the API key to remain private.

For the initial release of \mz, we distribute all mini-app packages and accompanying metadata in a single \texttt{tar.gz} archive, following the bulk-download model used for AndroZoo’s metadata (\url{https://androzoo.uni.lu/gp-metadata}). The archive is updated nightly, before 06:00 Luxembourg/Paris time, to include newly collected mini-app packages and metadata. As the collection grows, we plan to support individual package downloads using the package’s SHA-256 hash and a personal API key, following AndroZoo’s existing download model. 
\section{Research Opportunities}
\label{sec:opportunities}

A shared collection of mini-app packages and associated metadata can facilitate reproducible evaluations and systematic comparisons of existing research tools. For example, TaintMini~\cite{wang2023taintmini} and WeMinT~\cite{meng2023wemint}, which both use static taint analysis to track sensitive data flows, could be evaluated on the same set of packages from \mz. This would allow researchers to compare the identified flows, analysis coverage, and execution time under consistent experimental settings, reducing bias across independently collected datasets. More generally, \mz can facilitate further investigations in areas already explored by existing studies, including bug detection, data flows, privacy-policy consistency, credential leakage, and other topics~\cite{wang2022webug,wang2023taintmini,10.1145/3832140, meng2023wemint,wang2024minidetector,wang2025miniscope,ferrari2026telegapper,shi2025keymagnet,zhang2021measurement,li2024xposcope,li2024minitracker,wang2023onesize,wang2023apiscope,yang2022cmrf}.

\begin{figure}[ht]
\centering
\includegraphics[width=\columnwidth]{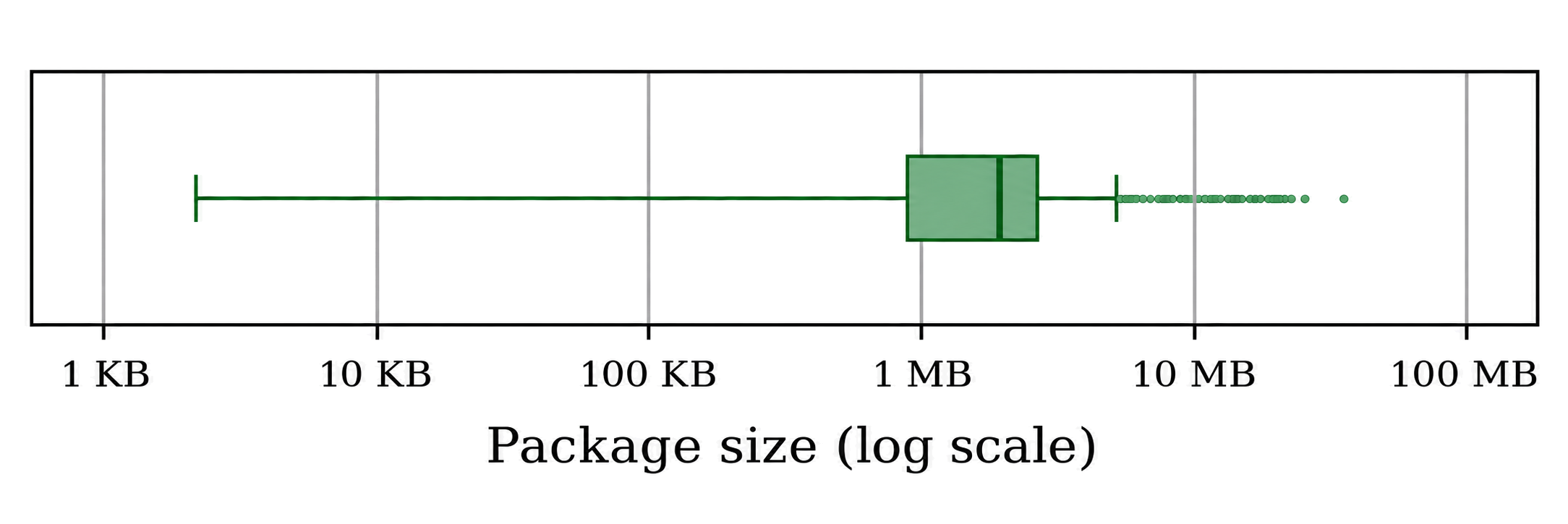}
\caption{Distribution of collected mini-app package sizes.}
\label{fig:miniapp-package-sizes}
\end{figure}

Beyond supporting existing lines of work, the continued collection of successive versions will enable longitudinal studies of mini-app evolution, including changes in package size, API usage, and dependencies. In the past, similar longitudinal studies, including analyses of multiple versions of the same apps, have been conducted on Android apps thanks to AndroZoo~\cite{calciati2018change,almanee2021quiet,he2018evolution,pendlebury2019tesseract,zhang2020apigraph,chen2023continuous}; such studies are now possible for mini apps as well through \mz. Additionally, beyond the packages themselves, \mz provides descriptions, developer information, version associations, and acquisition records that allow researchers to compare such characteristics or focus on specific mini-app groups. As an example, Figure~\ref{fig:miniapp-package-sizes} illustrates the size distribution of the collected mini-app packages.

\noindent
Although the median package size is \num{1.94}~MB, observed sizes span several orders of magnitude, from kilobytes to tens of megabytes. Researchers may use this information to compare mini apps with smaller and larger packages in terms of functionality, code reuse, or analysis cost, or to focus specifically on larger mini apps. These examples illustrate only some of the research opportunities offered by \mz; further studies could explore other combinations of package contents and metadata as the dataset evolves.
\section{Data Collection Challenges and Limitations}
\label{sec:limitations}
Collecting new mini apps and maintaining \mz involve challenges similar to those previously encountered with AndroZoo~\cite{allix2016androzoo,alecci2024retrospective}.

\noindent\textbf{Collection coverage.} As we cannot enumerate all available mini apps, we cannot precisely quantify the proportion of the ecosystem represented in \mz. Following AndroZoo’s approach, our objective is to collect and preserve as many mini apps as possible rather than guarantee exhaustive coverage.

\noindent\textbf{Temporal coverage.} Our crawlers collect different mini apps over time, so successive acquisitions of the same app do not necessarily capture consecutive releases. An app may receive multiple updates while the crawlers are collecting other mini apps, leaving gaps in its recorded version history.

\noindent\textbf{Collection maintenance.} Changes to the host app’s interface or APIs, access restrictions, and service unavailability may disrupt automated collection. Maintaining the dataset therefore requires monitoring collection failures and adapting the crawlers as the platform evolves.

\noindent\textbf{Platform scope.} The current collection covers only WeChat, whose collection workflow was the most straightforward to implement initially. Extending \mz to additional mini-app ecosystems is an explicit direction for future work.
\section{Conclusion}
\label{sec:conclusion}

We presented \mz, an evolving dataset of mini-app packages and associated metadata, together with its collection workflow and opportunities for reuse. 
Inspired by AndroZoo, our goal is to maintain a shared resource that reduces the acquisition effort required for mini-app research.
Future work will focus on expanding the collection through newly discovered mini apps, successive versions of previously collected apps, and additional \emph{super app} ecosystems, enabling longitudinal studies and comparisons across platforms. 
We hope that \mz will support ongoing research and encourage reproducible experiments within the mini-app research community.

\bibliographystyle{IEEEtran}
\bibliography{references}

\end{document}